\documentclass{article}
\PassOptionsToPackage{numbers,square,compress}{natbib}
\usepackage[dblblindworkshop,final]{neurips_2026}

\usepackage[utf8]{inputenc}
\usepackage[T1]{fontenc}
\usepackage{amsmath}
\usepackage{amssymb}
\usepackage{amsfonts}
\usepackage{bm}
\usepackage{booktabs}
\usepackage{caption}
\usepackage{float}
\usepackage{graphicx}
\usepackage{fontawesome5}
\usepackage{microtype}
\usepackage{natbib}
\usepackage{ragged2e}
\usepackage{subcaption}
\usepackage{xcolor}
\usepackage{hyperref}
\usepackage{url}
\workshoptitle{Sim2Science: ML with Imperfect Scientific Models}

\newcommand{\fig}[1]{Fig.~\ref{#1}}

\newcommand{\vit}{\texttt{ViViViT}}
\newcommand{\unet}{\texttt{U-Net}}

\definecolor{linkcolor}{rgb}{0.0,0.3,0.5}
\definecolor{darkred}{RGB}{175,0,0}
\definecolor{darkblue}{RGB}{14,0,185}
\definecolor{salmon}{RGB}{255,160,105}
\definecolor{darkteal}{RGB}{0,111,111}

\title{\vit{}: a Video-to-Video Vision Transformer for Cosmological Lightcone Reconstruction}

\author{%
  Federico Semenzato$^{1,2,3}$ \quad Alvise Raccanelli$^{1,2,4}$ \quad Michele Liguori$^{1,2,4}$ \\[4pt]
  \begin{minipage}{\dimexpr\textwidth-2\tabcolsep\relax}\centering\normalfont\small
    $^{1}$Dipartimento di Fisica e Astronomia ``G. Galilei'', Universit\`a degli Studi di Padova, Padova, Italy \\
    $^{2}$INFN, Sezione di Padova, Padova, Italy \\
    $^{3}$Department of Physics and Astronomy, Johns Hopkins University, Baltimore, MD 21218, USA \\
    $^{4}$INAF--Osservatorio Astronomico di Padova,
     Padova, Italy \\[3pt]
    \texttt{semenzato@jhu.edu} \\
    \texttt{alvise.raccanelli.1@unipd.it}
  \end{minipage}%
}

\begin{document}
\maketitle

\begin{abstract}
Moving along the line of sight of a cosmological lightcone also moves through cosmic time, as in a \textit{video} of cosmic structure traced by an evolving observable.
We recast the reconstruction of the underlying matter field from observed lightcones as a video-to-video translation task by introducing \vit, a vision transformer that factorizes attention across transverse frames and radial evolution.
On reconstructed matter fields from 21~cm neutral hydrogen simulations over~$z\in[7.66,16.00]$, \vit{} preserves spectral power and phase alignment within a few percent of the ground truth up to the Nyquist scale, and three-point correlations for all triangle configurations with sides up to~$k\simeq0.77\,h/\mathrm{Mpc}$, outperforming a convolutional 3D-\unet{} baseline across all measured statistics.
By encoding the evolving geometry of the cosmological observable itself, \texttt{ViViViT} marks a step toward field-level architectures that natively respect cosmic time, opening the path to time-aware reconstruction of cosmic tracers and targeted searches for redshift-dependent signatures in 21~cm data.

\end{abstract}

\begin{figure*}[t]
  \centering
  \begin{subfigure}[b]{0.28\textwidth}
    \centering
    \includegraphics[width=\textwidth]{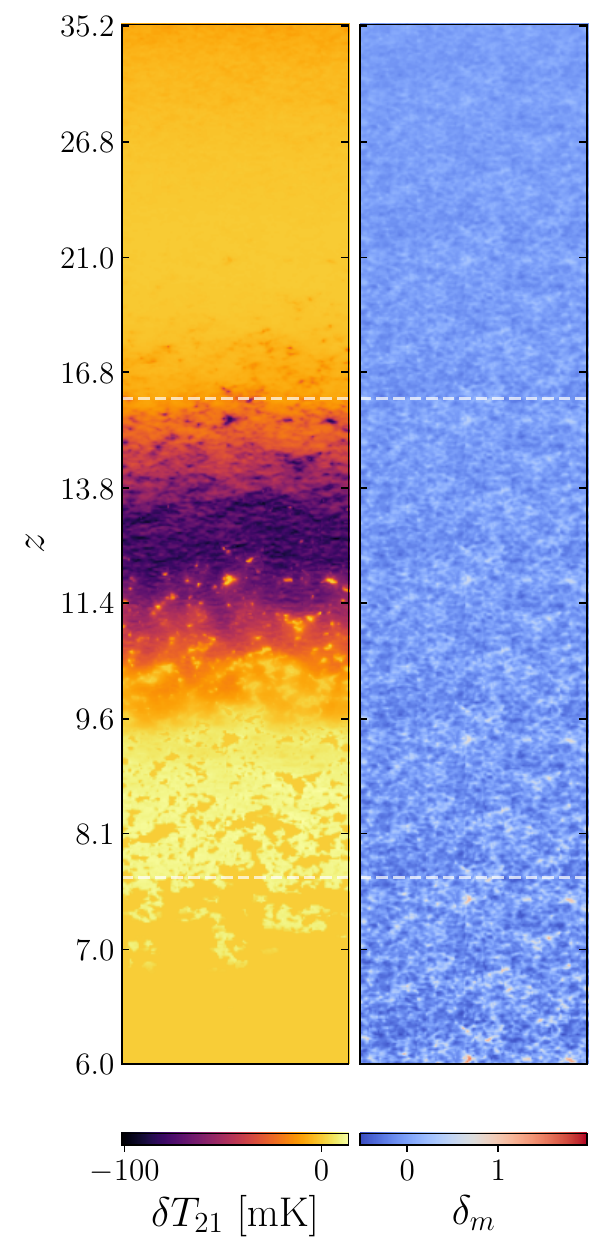}
    \caption{Evolving lightcone}
    \label{fig:overview_lightcone}
  \end{subfigure}
  \hfill
  \begin{subfigure}[b]{0.21\textwidth}
    \centering
    \includegraphics[trim={22.2cm 2cm 22.2cm 5cm},clip,width=\textwidth]{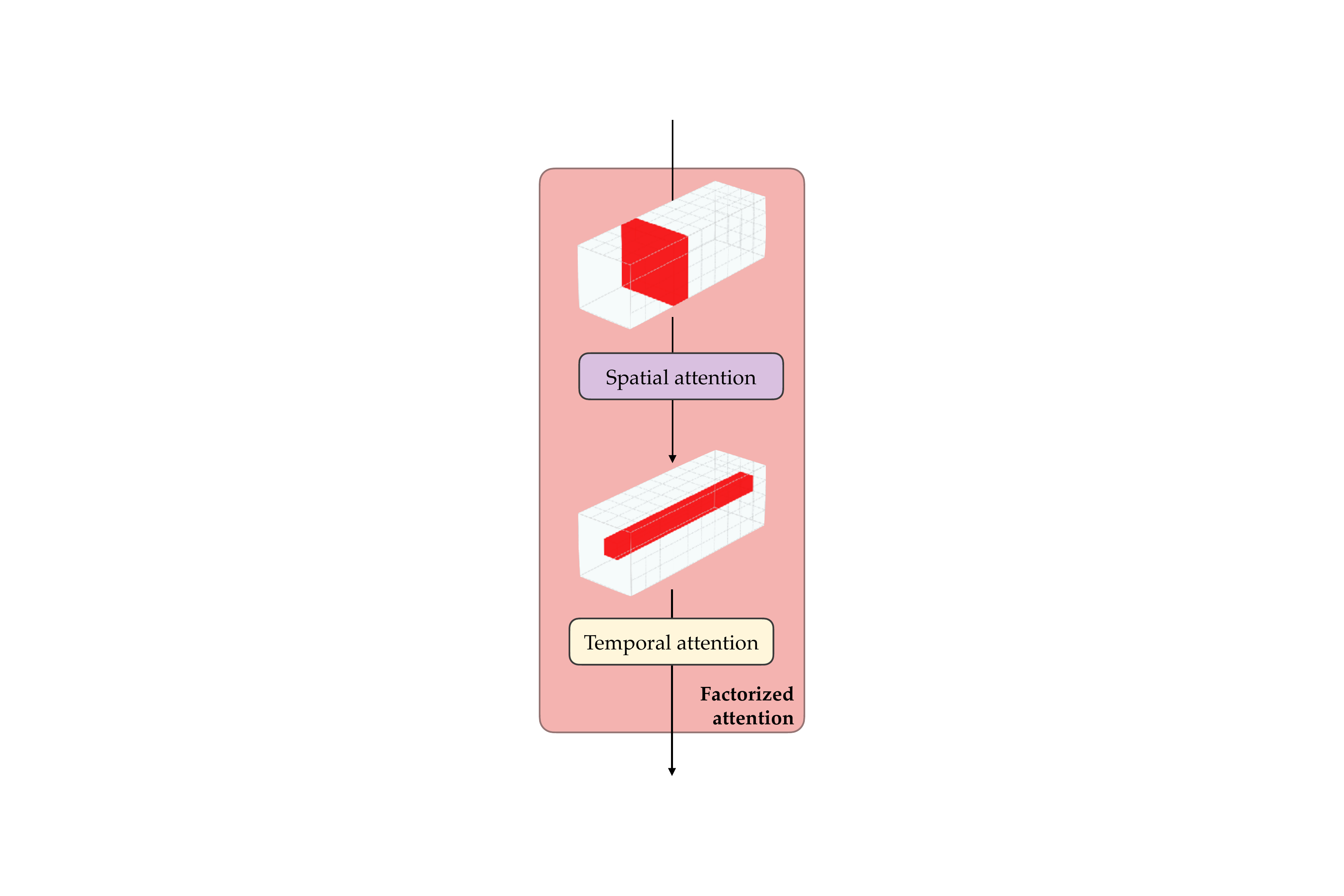}
    \caption{Factorized attention}
    \label{fig:overview_attention}
  \end{subfigure}
  \hfill
  \begin{subfigure}[b]{0.48\textwidth}
    \centering
    \includegraphics[trim={18cm 4.5cm 16.5cm 7cm},clip,width=\textwidth]{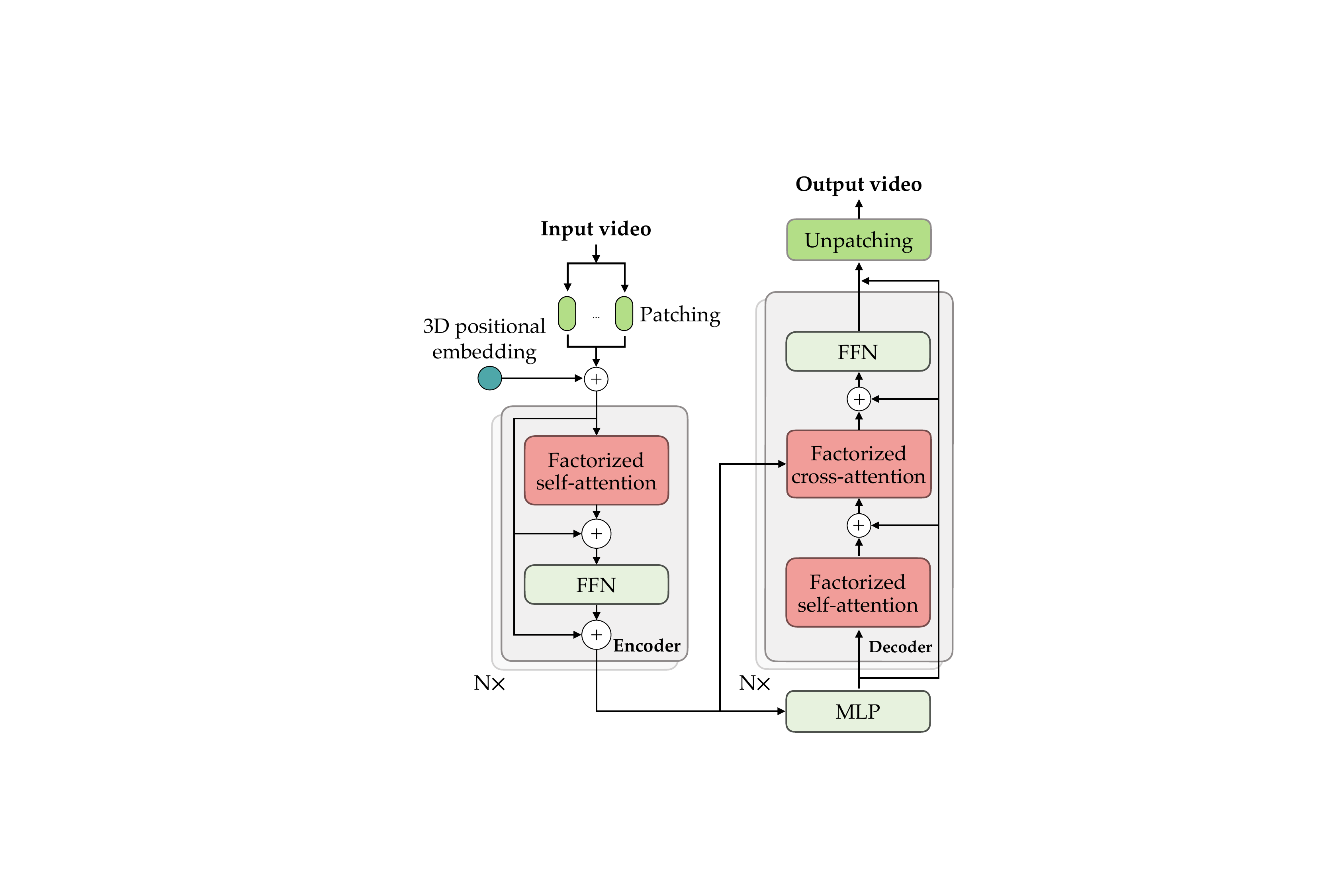}
    \caption{Field translation}
    \label{fig:overview_model}
  \end{subfigure}
  \caption{\justifying
  \textbf{A Video-to-Video Translation.}
  (a) Longitudinal temperature and density slices from simulated 21~cm lightcones.
  Redshift changes along the line of sight, with both brightness temperature~$\delta T_{21}$ and matter density~$\delta_m$ evolving across cosmic time.
  Dashed lines bound the reconstructed interval.
  (b) \vit{} alternates attention within transverse frames and along the line of sight.
  (c) An encoder--decoder translates the complete input sequence into a density sequence on the same grid.
  The decoder is initialized with a direct projection of the encoder's output.}
  \label{fig:overview}
  \vspace{-3mm}
\end{figure*}

\section{Introduction}
\label{sec:intro}

The redshifted 21~cm hyperfine transition of neutral hydrogen opens a direct, volumetric window onto the evolution of the Universe.
The 21~cm signal is a tracer of the underlying distribution of dark matter, following its gravitational growth across cosmic time.
Mapping its emission from resolved and unresolved galaxies and diffuse gas allows us to probe the local Universe, track how the first luminous sources heated and ionized the intergalactic medium, and potentially access the Dark Ages prior to the formation of the first stars (see, e.g.,~\cite{Furlanetto:2006jb, Pritchard:2011xb, Furlanetto:2015apc}).
The differential brightness temperature~$\delta T_{21}$ (relative to the CMB) is, however, not a direct image of the matter density~$\delta_m$: the signal is modulated  through the neutral hydrogen fraction, gas and spin temperatures, radiation backgrounds, and line-of-sight velocities.
The matter distribution shapes the signal through both the gas density and the abundance and clustering of galaxies, whose radiation heats and ionizes the surrounding intergalactic medium~\cite{Mesinger:2013nua}.
Exotic energy injection from sources such as decaying dark matter and primordial black holes can also alter the thermal history in a redshift-dependent way~\cite{Bernal:2017nec,Short:2019twc,Sun:2023acy}.
All these effects combine into a complex, non-linear mapping between the underlying matter field~$\delta_m$ and the observed 21~cm signal, with both local and time-varying contributions.

\paragraph{Reconstructing the matter distribution.}
Inverting this mapping would give direct access to the matter field shared by all cosmological tracers.
The evolution and statistical properties of~$\delta_m$ are governed by gravity, providing a more
robust and direct connection to early-Universe imprints~\cite{Munoz:2015eqa,deKruijf:2024voc}. Accessing this field would enable clean~$n$-point statistical measurements, cross-correlations with other tracers, and a powerful probe  to test the nature of cosmic acceleration, dark-matter physics, and departures from general relativity (see, e.g.,~\cite{Matarrese_2008,Guz+08,Ishak_2018,deKruijf:2024voc,DESI:2025zgx}).
Moreover, while depending on the astrophysical forward model, such a reconstruction would isolate how specific heating and ionization effects shape the signal, separating the coupled contributions of astrophysical processes from cosmic structure formation~\cite{Villanueva-Domingo:2020wpt}.

\paragraph{Towards tomographic 21~cm observations.}
Current radio interferometers (HERA~\cite{DeBoer:2016tnn}, LOFAR~\cite{vanHaarlem:2013dsa}, MWA~\cite{Tingay:2013oja}) target statistical detection of 21~cm power spectra at~$z \sim 8$--12~\cite{Parsons:2012ir, Dillon:2013dja}.
The Square Kilometre Array Observatory (SKAO) will extend the sensitivity to~$z \lesssim 28$ with baseline lengths enabling arcminute angular resolution, potentially achieving tomographic imaging of the neutral hydrogen structure~\cite{Koopmans:2015sua,Majumdar:2026dgt}.

\paragraph{Related work.}
Cosmological reconstruction has long inferred primordial initial conditions from late-time large-scale structure through Bayesian forward modeling~(e.g.,~\cite{Jasche:2012kq}), score-based posterior sampling~(e.g.,~\cite{Legin:2023ics}) and convolutional methods~(e.g.,~\cite{Floss:2023ylq,Chen:2023uup}). 
Those approaches connect a late-time field to a primordial snapshot, while our aim is to map between two evolving fields on the same lightcone.
Within 21~cm cosmology, existing reconstruction approaches largely inherit the inductive biases of convolutional networks, which have been used for parameter inference~\cite{Gillet:2018fgb,Hassan:2019cal}, matter-density reconstruction from 2D 21~cm slices~\cite{Villanueva-Domingo:2020wpt}, and for foreground and missing-mode mitigation~\cite{Makinen:2020gvh}. 
Generative models have also been used for foreground mitigation~\cite{sabti2024generativemodelingapproachreconstructing}.
A ViT--\unet{} model has been employed to emulate 21~cm volumes from matter and halo fields~\cite{Posture:2025cosmouit} and recently 
\citet{pietschke2026crosssimulatortransferfoundationmodel} introduced a self-supervised ViT encoder of 21~cm lightcones for cross-simulator transfer learning and parameter inference. 
Time dependence has been included as conditioning to CNNs to reconstruct the ionization fraction~\cite{Korber:2023pinion}, and in RNN architectures for parameter inference~\cite{Prelogovic:2021ljp}.
In this context, video models provide a mature basis to probe and generate spatiotemporal data, based on space--time transformers~\cite{Bertasius:2021vid,Arnab:2021vit} and diffusion systems~\cite{Ho:2022video,BarTal:2024lumiere,WanTeam:2025wan}.

\paragraph{Our contribution.}
The geometry of the lightcone implies that the radial axis is simultaneously comoving distance and look-back time: each transverse plane encodes a different epoch (and spatial location), so their sequence can be interpreted as a \textit{video} of cosmic evolution (\fig{fig:overview}, left), and reconstruction must account for correlations along that sequence.
We therefore recast 21~cm field reconstruction as video-to-video translation and develop \vit{}, a deterministic encoder--decoder transformer based on the ViViT architecture of~\cite{Arnab:2021vit}.
By factorizing attention into spatial and temporal components, our approach breaks the quadratic cost of full spacetime attention, allowing us to scale the model to larger volumes.
Moreover, by separating transverse and radial attention, we  provide a natural framework to capture signatures with different morphology and time dependence. 
We test the model on a matter reconstruction task and compare it to a convolutional \unet{} baseline.
Our results show that the transformer performs accurate field-to-field translation across the entire lightcone, improving  the reconstruction of the measured
two- and three-point statistics relative to the tested
convolutional baseline.

\section{Method}
\label{sec:method}

\subsection{21~cm physics and forward model}
\label{app:data}

The differential brightness temperature describes the redshifted 21~cm signal in emission or absorption relative to the CMB. Neglecting peculiar-velocity gradients, the optically thin signal from the diffuse intergalactic medium can be parameterized as
\begin{equation}
\delta T_{21} \approx 27 \, x_{\mathrm{HI}}(1+\delta_b) \left(1 - \frac{T_\gamma}{T_S}\right) \left(\frac{\Omega_b h^2}{0.023}\right) \left(\frac{0.15}{\Omega_m h^2}\right)^{1/2} \left(\frac{1+z}{10}\right)^{1/2} \, {\mathrm{mK}},
\label{eq:Tb_approx}
\end{equation}
where~$\delta_b$ is the baryon density contrast,~$x_{\mathrm{HI}} = 1 - x_{\mathrm{HII}}$ is the neutral hydrogen fraction (with~$x_{\mathrm{HII}}$ the ionized fraction),~$T_\gamma= 2.73(1 + z)$ K is the CMB temperature and~$T_S$ is the spin temperature, set by radiative coupling to the CMB, collisions and Wouthuysen--Field coupling to the Ly$\alpha$ background.
Peculiar velocities modify the line optical depth through line-of-sight velocity gradients and shift the correspondence between comoving position and observed frequency~\cite{Furlanetto:2006jb,Pritchard:2011xb}. In standard thermal history models, the gas cools adiabatically after thermally decoupling from the CMB.
While collisions remain effective, they couple $T_S$ toward the gas kinetic temperature $T_K<T_\gamma$, so that the signal appears in absorption. 
As the Universe expands and gas density drops, collisional coupling becomes highly inefficient and radiative coupling to the CMB then dominates, pulling $T_S$ toward $T_\gamma$, suppressing the diffuse 21~cm signal.
During cosmic dawn, Ly$\alpha$ photons from the first stars introduce Wouthuysen-Field coupling, driving $T_S$ toward the Ly$\alpha$ colour temperature, which is similar to $T_K$.
This produces absorption, if the gas remains colder than the CMB.
X-ray heating raises $T_K > T_\gamma$; with effective spin coupling, the signal then appears in emission ($\delta T_{21} > 0$).
During patchy reionization, ionized regions locally suppress the diffuse 21~cm signal.
Fluctuations in $x_{\mathrm{HI}}$ can dominate the brightness temperature structure when spin temperature fluctuations become subdominant. Temporal structure of different stages depends on the source population, possible exotic sources, and their radiative properties.

We generate paired temperature and density lightcones using \texttt{21cmFAST}~\cite{Mesinger:2010ne, Munoz:2021psm}, a simulation framework which combines perturbation theory for the density field with semi-numerical prescriptions for the ionization and thermal state of the intergalactic medium, to produce large-volume realizations of the 21~cm signal.
Our task maps~$\delta T_{21}(\mathbf{x}_\perp,z)$ to~$\delta_m(\mathbf{x}_\perp,z)$ on a volume~$\mathcal V\in\mathbb R^{T\times H\times W}$ spanning~$z\in[7.66,16.00]$, with~$(T,H,W)=(384,128,128)$ voxels of comoving size~$4\,\mathrm{Mpc}$. We choose this interval to include morphologically rich structure across absorption, heating and patchy reionization, testing reconstruction as the temperature--density relation evolves on a fixed grid. 
Each channel uses one global mean and standard deviation; we do not normalize individual slices, so the model must learn the evolution of the field amplitude along the lightcone.

The cosmological parameters are fixed to the Planck 2018 values:~$h = 0.6727$,~$\Omega_m = 0.3166$,~$\Omega_b = 0.0494$,~$n_s = 0.9649$, and~$\sigma_8 = 0.8120$~\cite{Planck:2018vyg}.
For the astrophysical parameters, we employ the parameterization of~\cite{Park:2018ljd}, with prior ranges following~\cite{sabti2024generativemodelingapproachreconstructing}:
the star formation efficiency at~$M_h = 10^{10} M_\odot$ is sampled from~$\log_{10} f_{\star,10} \sim \mathcal{U}(-1.45, -1.15)$, the ionizing photon escape fraction from~$\log_{10} f_{\mathrm{esc},10} \sim \mathcal{U}(-1.17, -0.84)$, and the X-ray luminosity per unit star formation rate from~$\log_{10}(L_X / {\mathrm{SFR}}) \sim \mathcal{U}(40.45, 40.55)$ in units of~$\log_{10}(\mathrm{erg~s}^{-1}~\mathrm{yr}~M_\odot^{-1})$.
Mass dependence of these quantities employs power-law scaling with indices~$\alpha_\star = 0.5$ and~$\alpha_{\mathrm{esc}} = -0.5$.
From each simulation run, we extract two volumetric fields: differential brightness temperature~$\delta T_{21}(\mathbf{x}_\perp, z)$ encoding the observable 21~cm signal and matter density contrast~$\delta_m(\mathbf{x}_\perp, z)$.
With multiple output channels, the model could jointly reconstruct matter density and the neutral fraction~$x_{\mathrm{HI}}(\mathbf{x}_\perp,z)$, including its spatial topology.
We train on 3000 realizations with a 90/10 train--validation split and reserve a separate additional set of 200 lightcones for all reported statistics.

\texttt{21cmFAST} evolves the field in coeval cubic boxes of side~$L_{\rm box}=512$ Mpc, periodic on all three Cartesian axes, at a discrete set of node redshifts. It then connects the lightcone by extracting and interpolating transverse slices from the boxes whose redshifts bracket each radial pixel~\cite{Mesinger:2010ne, Munoz:2021psm}.
The temperature--density relation hence evolves, but periodicity imprints from the same comoving structure enter the simulated lightcone at different times.  
This construction differs from an observed lightcone, where separated radial intervals occupy distinct comoving regions.
A model with global line-of-sight reach can use the repeated structure imprints, while one with a shorter receptive field cannot.
In Appendix~\ref{app:toy}, we test our model in a setting without any simulator artifacts to isolate the receptive-field advantage.

\subsection{Video-to-video transformer}
\label{sec:model}
We build an encoder--decoder transformer that translates an input temperature lightcone into a density lightcone on the same grid.
In this work, we focus on a one-channel volume. Additional input channels could combine the 21~cm lightcone with co-spatial galaxy or line-intensity maps, while multiple output channels could jointly reconstruct density and neutral fraction.
Before encoding, a strided 3D convolution forms overlapping spatial patches, which mitigates high-frequency discontinuities at patch boundaries.
Full attention over all tokens would be computationally impractical; therefore, we factorize attention into alternating spatial and temporal components (see central panel of \fig{fig:overview}), reducing the cost from~$\mathcal O(N_t^2N_s^2)$ to~$\mathcal O[N_tN_s(N_t+N_s)]$, where~$N_s=N_hN_w$~\cite{Arnab:2021vit}.
We then project each patch to a~$d$-dimensional embedding space, including a learned factorized positional encoding to retain spatial and temporal information (see Appendix~\ref{app:architecture} for details).

As visually summarized in the right panel of \fig{fig:overview}, the encoder forms a global contextual representation of the input tokens.
A feed-forward projection initializes the decoder from the encoder output.
This initialization preserves the token-wise correspondence between the input and target lightcones while passing the encoder's spatial and radial context to the decoder.
The decoder uses self- and cross-attention, followed by a local convolutional upsampling head that restores the full grid.
Temporal self-attention is lower-triangular at patch resolution.
Appendix~\ref{app:architecture} provides the full architecture and training details.

We use a pixel-wise~$L_1$ loss that penalizes absolute deviations between predicted and target fields, combined with a gradient-difference loss that enforces similarity in local structure and edge sharpness.
The complete objective and a controlled~$L_1$--$L_2$ ablation are reported in Appendix~\ref{app:loss}.

\section{Experiments}
\label{sec:results}

We compare \vit{} with a four-level 3D \unet{}.
In Appendix~\ref{app:toy}, we first compare the temporal receptive fields of the two architectures on a synthetic benchmark, varying the \unet{} depth: \vit{} retains global context across the entire temporal direction, while the \unet{} loses long-range accuracy as the target correlation length grows.
For our main cosmological task, we train both models on the same 3000 lightcones and evaluate them on a held-out set of 200 lightcones by measuring the two-point and three-point statistics of the reconstructed density fields, which probe two-point correlations and non-Gaussian mode coupling relevant for cosmological inference~\cite{Matarrese:1997sk,Eisenstein_1998,Watkinson:2017kfg,Coulton:2022qbc}.
The code and trained checkpoints are available \href{https://github.com/fsemenzato/ViViViT.git}{\faGithub{}}.

\begin{figure}
\centering
\includegraphics[width=0.65\textwidth]{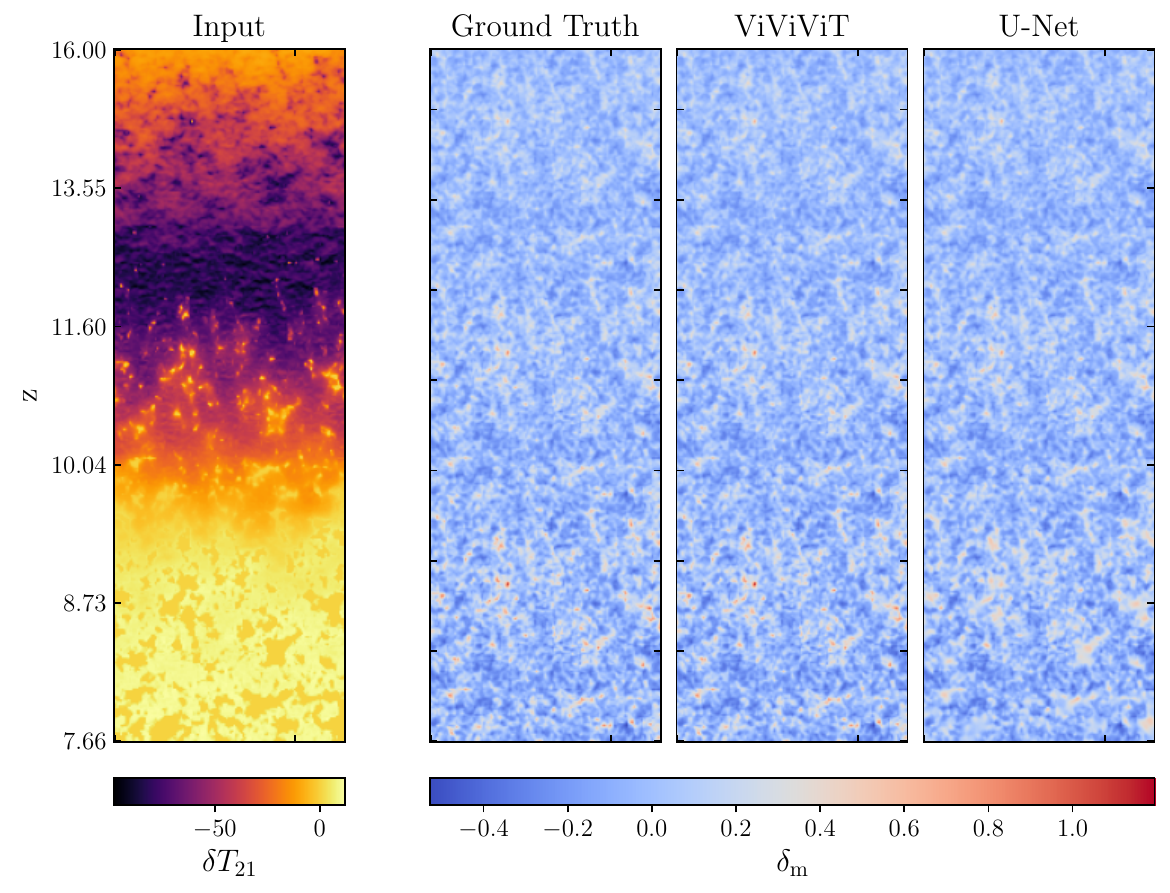}
\caption{\justifying
\textbf{\vit{} retains finer density structure than the tested \unet{} on a held-out longitudinal slice.}
Input 21~cm brightness temperature~$\delta T_{21}$ (in mK), target matter density, and both reconstructions share a color scale within each field type.}
\label{fig:field_slices}
\end{figure}

\begin{figure}[t]
  \centering
  \includegraphics[width=\columnwidth]{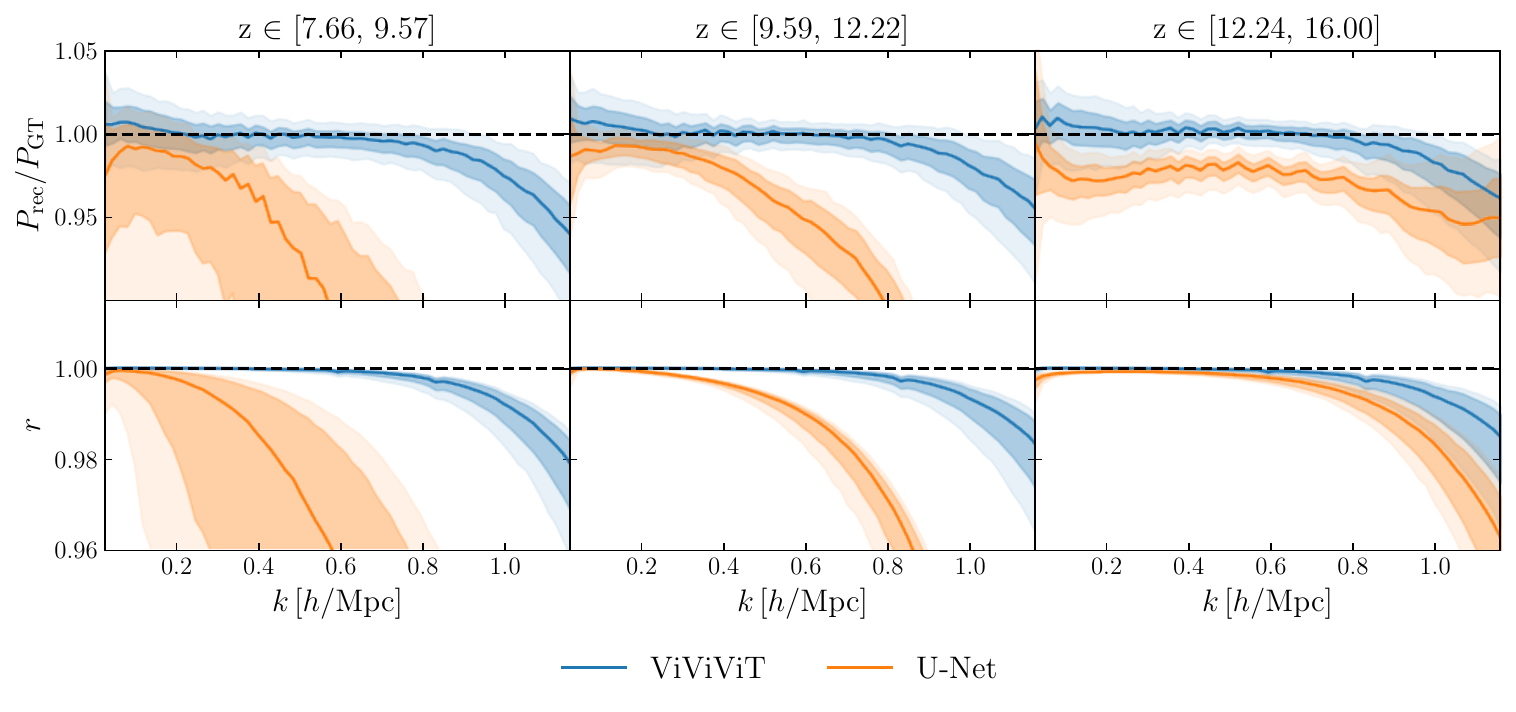}
  \vspace{1mm}
  \includegraphics[width=\columnwidth]{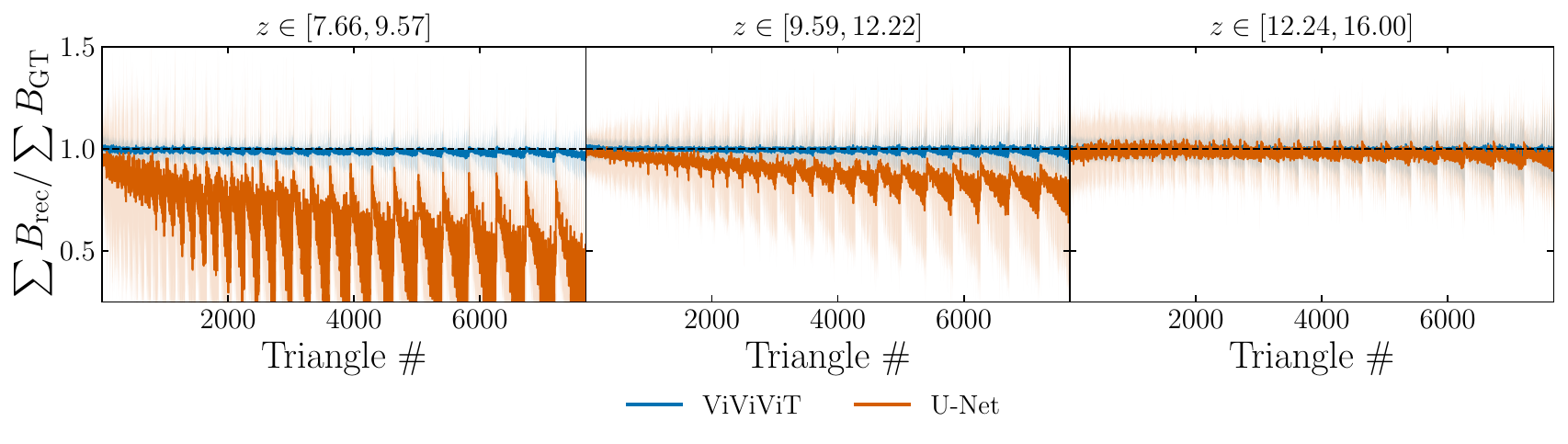}
  \caption{\justifying
  \textbf{\vit{} recovers the measured two- and three-point statistics more accurately than the tested \unet{}.}
  \emph{Top:} Reconstructed-to-target power spectrum ratio and cross-correlation coefficient~$r(k)$ in three consecutive redshift bins; 
  \emph{Bottom:} reconstructed-to-target isotropic bispectrum ratio for all retained closed triangles, ordered by their largest mode and shape.
  Curves are medians over 200 held-out lightcones and bands span the 16th--84th and 2.5th--97.5th percentiles.
  \vit{} (blue) stays closer to unit power and phase coherence at large~$k$ and preserves most three-point configurations, while the \unet{} (orange) develops a scale- and redshift-dependent deficit.}
  \label{fig:statistics}
\end{figure}

\paragraph{Field morphology.}
\fig{fig:field_slices} presents a qualitative comparison of the reconstructed density fields~$\hat{\delta}_m$ with the ground truth~$\delta_m$, along a longitudinal slice through the lightcone.
The input 21~cm brightness temperature (left panel) exhibits the characteristic evolution of the signal: absorption deepening toward $z \approx 12$--$13.5$, followed below $z \approx 12$ by heated and ionized regions that grow until ionized bubbles with $\delta T_{21} \approx 0$ dominate at $z \lesssim 10$.
The reconstruction task requires inverting this non-monotonic mapping while disentangling astrophysical evolution from gravitational clustering.
Both architectures successfully recover the large-scale cosmic web, correctly placing major overdensities and underdensities.
The architectures differ in the morphological details they recover.
The \unet{} baseline tends to produce overly smooth fields, blurring the sharp contrast between filaments and voids.
In contrast, the \texttt{ViViViT} model retains sharper features and better preserves the texture of the density field.
Crucially, the reconstruction remains stable also at low redshift ($z \lesssim 10$), where the 21~cm signal is highly non-linear and dominated by ionized regions with near-zero brightness temperature, indicating that the models successfully adapt to the evolving relationship between~$\delta T_{21}$ and~$\delta_m$.

\paragraph{Two-point fidelity.}
We divide each reconstructed lightcone into three adjacent~$128^3$ regions and measure power after tapering only the non-periodic radial direction (see Appendix~\ref{app:estimators} for details on the estimator).
The power spectrum~$P(k)$ measures the variance of Fourier modes as a function of scale, whereas the cross-correlation coefficient~$r(k)$ measures phase alignment between the reconstructed and target fields (see Appendix~\ref{app:estimators} for definitions).
In \fig{fig:statistics}, the \unet{} loses small-scale power earliest in the lowest-redshift bin, where growing ionized regions locally erase the temperature signal.
\vit{} keeps the median power ratio near unity over most measured modes.
More importantly, it maintains cross-correlation coherence~$r(k)>0.99$ to~$k\simeq0.9\,h/\mathrm{Mpc}$ in the central bin, while the tested \unet{} crosses that threshold near~$0.6\,h/\mathrm{Mpc}$.
These diagnostics show that \vit{} improves both
small-scale power recovery and cross-correlation with the target
field relative to the tested \unet{}.
The longitudinal slices in \fig{fig:field_slices} illustrate
the corresponding differences in reconstructed morphology.

\paragraph{Non-Gaussian fidelity.}
Non-linear gravitational collapse couples Fourier modes of~$\delta_m$, so we next measure the bispectrum~$B(k_1,k_2,k_3)$ to quantify whether the non-Gaussian structure is preserved (see Appendix~\ref{app:estimators} for the estimator construction).
\vit{} reconstructs the ground-truth bispectrum to within about 5\% for most retained triangle configurations, while the \unet{} reconstruction error grows toward smaller scales and lower redshift (bottom panel of \fig{fig:statistics}).
Once local brightness temperature is suppressed inside an ionized region, recovering the surrounding density requires context from outside that region.

\begin{figure}
\centering
\includegraphics[width=\textwidth]{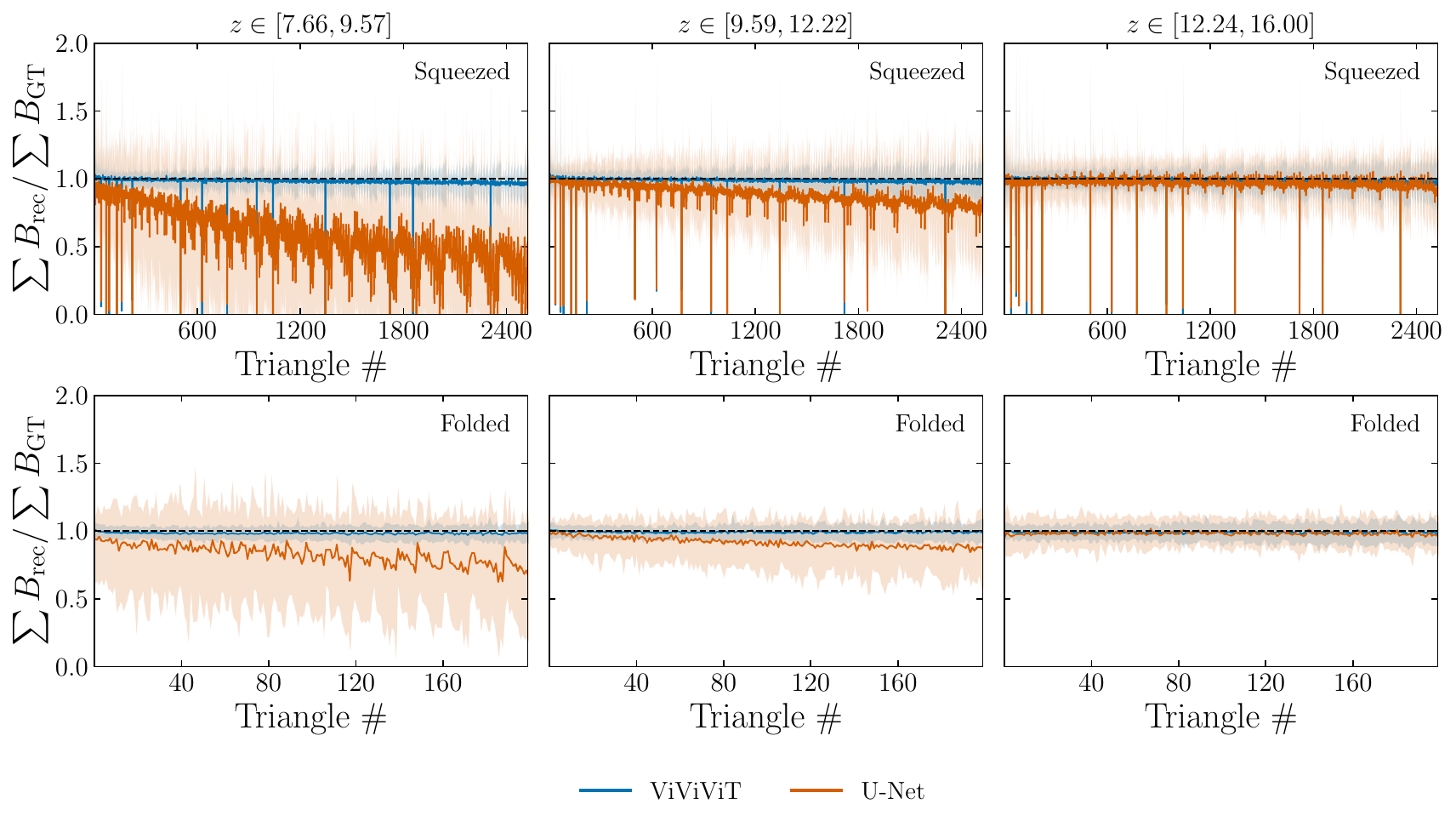}
\caption{\justifying
\textbf{\vit{} retains line-of-sight-aligned three-point structure more accurately than the tested \unet{}.}
Reconstructed-to-target bispectrum ratios for squeezed (top) and folded (bottom) triangle families across the three redshift bins.
Curves are medians over the held-out set and bands span the 16th--84th percentiles.
Sharp excursions occur where the target bispectrum approaches zero and the ratio becomes ill-conditioned; they should not be interpreted as enhanced reconstruction signal.
The plotted ratios and percentile bands are clipped to the displayed vertical range.}
\label{fig:los_bispectrum}
\end{figure}

\paragraph{Line-of-sight-aligned three-point structure.}
The isotropic result in \fig{fig:statistics} averages over triangle orientation.
To target the configurations that encode long-range correlations along the line of sight, we bin each bispectrum mode into five~$\mu = k_\parallel/k$ bins. 
We then isolate the bispectrum configurations aligned along the line of sight by keeping only triangles whose long mode falls in the line-of-sight-aligned angular bin ($\mu_{\mathrm{long}} \equiv k_{\parallel,\mathrm{long}}/k_{\mathrm{long}} \ge 0.8$).
This selection fixes the long radial mode while leaving the transverse shape free, and \fig{fig:los_bispectrum} shows the resulting amplitude ratio for the squeezed ($k_2 \approx k_1 \gg k_3$, in practice~$k_3/k_1 \le 0.5$ and~$k_2/k_1 \ge 0.8$) and folded ($k_1/(k_2 + k_3) \ge 0.85$ and~$k_3/k_1 > 0.5$) families.
\vit{} better preserves three-point amplitudes for triangles whose longest-wavelength leg is aligned with the line of sight.
The \unet{} loses the radial long-mode power progressively with decreasing redshift: the squeezed ratio falls from near unity in the farthest redshift bin to~$\sim 0.5$ at the smallest scales of the closest one, and the folded family follows the same trend more mildly.
The \unet{} architecture does not preserve accurate spectral information when processing the lightcone section corresponding to the epoch of reionization, failing to fully recover structure information from the geometry of ionized regions.

\section{Conclusions}
\label{sec:conclusions}
A cosmological lightcone can be viewed as a video across cosmic time, with successive transverse slices encoding successive epochs of structure formation at distinct spatial regions.
We have introduced \texttt{ViViViT}, a spacetime-aware transformer that processes the lightcone through factorized attention, learning a one-shot translation from one cosmological field to another.
On held-out \texttt{21cmFAST} lightcones, we showed that \vit{} outperforms the tested convolutional baseline in preserving the measured two- and three-point statistics of the target field.
A separate  synthetic benchmark demonstrates its ability to exploit long-range radial information when the reconstruction task requires it.

\paragraph{Limitations.}
As discussed in Section~\ref{app:data}, the \texttt{21cmFAST} forward model introduces periodicity artifacts from the same comoving structure at different times, differently from a real observed lightcone and that may provide additional reconstruction information.  The synthetic benchmark (see Appendix~\ref{app:toy}) shows that \texttt{ViViViT} can exploit long-range radial information when it is physically present.
This benchmark establishes long-context capability, but does not isolate its contribution to the cosmological task. We aim to extend our method to more refined lightcone simulations and further recast the field translation task as biased tracer emulation from matter density to observed lightcone.
Our mapping is a deterministic reconstruction: the model produces the entire volume in a single forward pass, but it does not preserve posterior uncertainty.
In future work, our aim is to adapt this infrastructure to conditional sampling, accessing the conditional distribution rather than its median.
Moreover, applications to real datasets must include observational and astrophysical systematics and project the physical field to an instrumental measurement.
In particular, the galactic and extragalactic foregrounds confine reliable spectral information outside a wedge-shaped region in Fourier space.
Neural approaches to foreground mitigation~\cite{Makinen:2020gvh,sabti2024generativemodelingapproachreconstructing} could be combined with our framework, and the effectively global receptive field of factorized attention is well suited to inpainting the lost modes from the surviving, correlated ones outside the wedge.
A complementary route against both the wedge and the intrinsic information loss inside ionized regions is to feed the network several tracers of the same underlying mass field: \texttt{ViViViT} extends naturally to multi-channel input, exploiting cross-correlations between tracers to potentially break the foreground degeneracy.
We leave a detailed study of these systematics and their mitigation for future work.

Beyond the present reconstruction, a natural possible extension is to target redshift-dependent physical processes whose signature is distributed across the lightcone.
Exotic energy injection from decaying or annihilating dark matter and from primordial black holes can modify the thermal and ionization history of the intergalactic medium~\cite{Bernal:2017nec,Short:2019twc,Sun:2023acy,Sun:2025dmpbh,deKruijf:PBHDarkAgesInPrep}, while modified growth histories alter the evolution of the matter field.
These effects are typically small in amplitude and easily confused with astrophysical uncertainties in power-spectrum analyses, yet coherent across the lightcone.

To our knowledge, \texttt{ViViViT} is the first video-to-video transformer applied to cosmological matter-field lightcone reconstruction. More broadly, it provides a flexible framework that directly encodes the lightcone spacetime structure, enabling more robust extraction of physical information from these complex datasets.

\begin{ack}
The authors thank Nicola Bellomo, Jessie de Kruijf, Sarah Libanore, Phoenix, Francesco Spezzati and Benjamin Wandelt for insightful feedback and discussions.
The authors also thank the anonymous reviewers for their constructive feedback, which was useful in refining the presentation of our work.
This work is partly supported by ICSC - Centro Nazionale di Ricerca in High Performance Computing, Big Data and Quantum Computing, funded by European Union - NextGenerationEU.
ML acknowledges support by the MIUR Progetti di Ricerca di Rilevante Interesse Nazionale (PRIN) Bando 2022 - grant 20228RMX4A.
AR acknowledges funding from the Italian Ministry of University and Research (MIUR) through the ``Dipartimenti di eccellenza'' project ``Science of the Universe''.

\end{ack}

\section*{LLM Usage}
In preparing this paper, we used large language models as assistive tools. In particular, LLMs were used for editing the draft for clarity, distributed training setup, running numerical experiments, and polishing figures.
The authors take full responsibility for the content of this paper.

\appendix

\section{Architecture and training details}
\label{app:architecture}
Tables~\ref{tab:vivit_architecture} and~\ref{tab:unet_architecture} report the two model specifications; Table~\ref{tab:training} lists the optimization settings shared by both architectures.
On four A100 GPUs, one complete training-and-validation epoch takes $\sim 52$ minutes for \vit{} and $\sim 24$ minutes for \unet{}. We extracted minimum-validation-loss trained checkpoints after 260 and 409 epochs, respectively.

\paragraph{Tokenization and position.}
The tokenizer applies a 3D convolution with kernel~$(p_t,p_s+2,p_s+2)$ and stride~$(p_t,p_s,p_s)$, where~$p_t$ denotes the temporal patch size and~$p_h = p_w \equiv p_s$ the spatial patch size in voxel units.
Spatial neighbors therefore overlap by two voxels, whereas the radial patches remain aligned and non-overlapping.
We use this strided convolution to avoid high-frequency discontinuities at patch boundaries, which manifest as spurious power in the Fourier spectrum at wavenumbers~$k \sim 2\pi/p_s$ where~$p_s$ is the spatial patch size, converted to physical units.
The tokenizer maps the volume to~$\mathbf Z^{(0)}\in\mathbb R^{B\times N_t\times N_s\times d}$, where~$B$ is the batch size, 
$N_t$ is the number of temporal patches,~$N_s$ is the number of spatial patches, and~$d$ is the embedding dimension.
We also add factorized learned position vectors,
\begin{equation}
\mathbf P_{i,j,k}=\mathbf P_i^{(t)}+\mathbf P_j^{(h)}+\mathbf P_k^{(w)}\,,
\end{equation}
where~$\mathbf{P}^{(t)} \in \mathbb{R}^{N_t \times d}$,~$\mathbf{P}^{(h)} \in \mathbb{R}^{N_h \times d}$, and~$\mathbf{P}^{(w)} \in \mathbb{R}^{N_w \times d}$ are independent learnable parameters initialized with small random values. This additive decomposition provides unique encodings while using~$\mathcal O(N_t+N_h+N_w)$ positional parameters instead of a table over the full token grid.
The input lightcone is thus represented as a sequence of tokens with embedded positional information.

\paragraph{Attention and decoding.}
Each block applies spatial attention independently within every radial patch plane, followed by temporal attention independently at every spatial token.
Temporal self-attention uses lower-triangular masks, while temporal cross-attention to the encoder is unmasked.
The decoder is initialized with a direct projection of the encoder's output via a non-linear MLP, and decoder blocks combine masked self-attention with encoder cross-attention.
A local convolutional head mixes the coarse token grid and restores resolution through two joint~$2\times2\times2$ sub-pixel rearrangement stages.
Token representations are first projected from embedding dimension~$d$ to an internal feature dimension~$d_{\text{feat}} = 256$.
The resulting features are reshaped to form a 3D grid matching patch resolution:~$(B, d_{\text{feat}}, N_t, N_h, N_w)$.
Before upsampling, this coarse feature grid undergoes mixing through circular convolutions, distributing information across tile boundaries before upsampling to avoid the magnification of potential artifacts.
Resolution is recovered through two stages, each applying~$2\times$ upsampling jointly along all three axes by employing pixel shuffle operations followed by residual refinement.
The transformation is a deterministic tensor rearrangement, requiring no interpolation or learned kernels.
This approach, first introduced in super-resolution tasks ~\cite{2016arXiv160905158S}, effectively separates the upsampling (rearrangement) from the feature generation (convolution), preventing artifacts from propagating both spatially and in Fourier space.
The final layers apply residual refinement and project to the output channel dimension through additional circular convolutions, producing the reconstructed field at full resolution~$(T, H, W)$.

\noindent
\begin{minipage}[t]{0.49\textwidth}
\centering
\small
\captionsetup{hypcap=false}
\captionof{table}{\vit{}}
\label{tab:vivit_architecture}
\begin{tabular}{@{}ll@{}}
\toprule
Patch kernel / stride & $(4,6,6)$ / $(4,4,4)$ \\
Token grid & $96\times32\times32$ \\
Embedding dimension & $d=512$ \\
Attention heads & 8 \\
Encoder blocks & 6 \\
Decoder blocks & 6 \\
Feed-forward dimension & 2048 \\
Decoder initialization & MLP projection \\
Upsampling feature width & 256 \\
Dropout & 0.1 \\
\bottomrule
\end{tabular}
\end{minipage}\hfill
\begin{minipage}[t]{0.49\textwidth}
\centering
\small
\captionsetup{hypcap=false}
\captionof{table}{3D \unet{}}
\label{tab:unet_architecture}
\begin{tabular}{@{}ll@{}}
\toprule
Convolution kernel & $3\times3\times3$ \\
Initial feature width & 32 \\
Encoder widths & 32, 64, 128, 256 \\
Bottleneck width & 512 \\
Decoder widths & 256, 128, 64, 32 \\
Downsampling levels & 4 \\
Upsampling levels & 4 \\
Bottleneck grid & $24\times8\times8$ \\
\bottomrule
\end{tabular}
\end{minipage}

\begin{table}[H]
\centering
\small
\caption{Shared optimization settings for the cosmological task.}
\label{tab:training}
\begin{tabular}{ll}
\toprule
Loss & $L_1$ pixel $+L_1$ gradient difference \\
Optimizer & AdamW, $(\beta_1,\beta_2)=(0.9,0.95)$, weight decay 0.05 \\
Learning rate & $10^{-4}$ with warmup and cosine decay \\
Gradient clipping & norm 1.0 \\
Batch / accumulation / GPUs & 1 / 4 / 4 \\
Maximum epochs / patience & 500 / 30 \\
Precision & automatic mixed precision \\
\bottomrule
\end{tabular}
\end{table}

\section{Training objective and loss ablation}
\label{app:loss}

\begin{figure}
\centering
\includegraphics[width=0.85\textwidth]{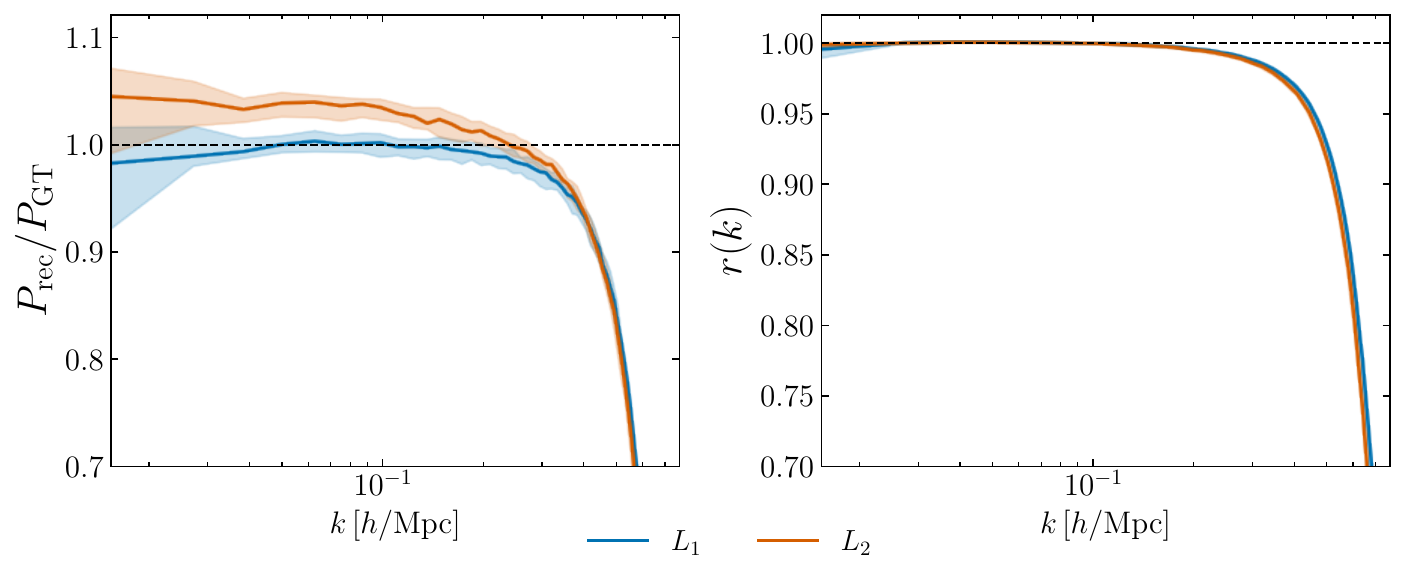}
\caption{\justifying
\textbf{Loss-norm ablation for~$\delta T_{21}\rightarrow\delta_m$}.
Lines show medians over 27 held-out volumes; bands span the 16th--84th percentiles.
The two objectives are largely consistent across the measured scales, while the power spectrum ratio mildly favors the~$L_1$ training objective.}
\label{fig:loss_ablation}
\end{figure}

For target~$y$ and reconstruction~$\hat y$ over volume~$\mathcal V$, the training objective is
\begin{align}
\mathcal L_{\rm pix}
  &=\frac{1}{|\mathcal V|}\sum_{\mathbf x\in\mathcal V}
    |\hat y(\mathbf x)-y(\mathbf x)|,\\
\mathcal L_{\rm gdl}
  &=\frac{1}{3|\mathcal V|}\sum_{d\in\{t,h,w\}}
    \sum_{\mathbf x\in\mathcal V}
    |\nabla_d\hat y(\mathbf x)-\nabla_d y(\mathbf x)|,\\
\mathcal L&=\mathcal L_{\rm pix}+\lambda_{\rm gdl}\mathcal L_{\rm gdl},
\qquad \lambda_{\rm gdl}=1.
\end{align}
The~$L_1$ pixel term measures voxel-wise fidelity without quadratically upweighting locally large residuals, which can arise where the 21 cm signal weakly constrains the density; the directional term instead penalizes mismatches in local field variations.
We isolate the norm choice on a reduced~$128^3$ task with 500 simulations and the same 80-epoch budget for~$L_1$ and~$L_2$ versions of both terms.
\fig{fig:loss_ablation} shows comp atible large-scale performance with a modest advantage for~$L_1$ at large scales.

\section{Summary-statistic estimators}
\label{app:estimators}

We split each~$384\times128^2$ reconstructed volume into three contiguous~$128^3$ regions.
The transverse axes inherit periodic simulation boundaries, but the radial field evolves and is not periodic within a redshift bin, requiring radial tapering for proper power spectrum and bispectrum estimation.

Before transforming to Fourier space, we then multiply both reconstruction and target ground-truth fields by the same one-dimensional cosine-tapered window along the radial axis, where~$t$ represents the radial frame index within a given redshift bin, with~$N=128$ frames per bin,
\begin{equation}
w(t)=
\begin{cases}
\frac12\!\left[1+\cos\!\left(\pi\!\left(\frac{2t}{\alpha(N-1)}-1\right)\right)\right],
& t<\frac{\alpha}{2}(N-1),\\[4pt]
1, & \frac{\alpha}{2}(N-1)\le t\le\left(1-\frac{\alpha}{2}\right)(N-1),\\[4pt]
w(N-1-t), & \text{otherwise},
\end{cases}
\label{eq:tukey}
\end{equation}
with taper fraction~$\alpha=0.5$, i.e.\ unity over the central half of the region and half-cosine ramps over the outer quarters.
Each region is first reduced to a fluctuation field by subtracting its mean, and the tapered field
\begin{equation}
\delta_w(\mathbf x_\perp,t)=w(t)\left[\delta(\mathbf x_\perp,t)-\bar\delta_w\right],
\qquad
\bar\delta_w=\frac{\sum_{\mathbf x}w(t)\,\delta(\mathbf x)}{\sum_{\mathbf x}w(t)},
\label{eq:taperfield}
\end{equation}
is exactly zero-mean by construction.
The window multiplies ground truth and reconstruction identically, modifying the measured spectra of both fields.
The window attenuation is compensated by the moments~$W_n=N^{-1}\sum_t w(t)^n$, entering the estimators below.
No mass-assignment deconvolution is applied because the fields are defined directly on the grid rather than painted from particle simulations.

\paragraph{Power spectrum.}
For two tapered fields~$\delta_w,\delta'_w$ on the~$N^3$ grid  we estimate
\begin{equation}
\widehat P_{\delta\delta'}(k_b)=\frac{V}{N^6\,W_2}\frac{1}{N_b}
\sum_{\mathbf k\in \mathcal S_b}\mathrm{Re}\!\left[\widehat\delta_w(\mathbf k)\,\widehat\delta'^{\,*}_w(\mathbf k)\right],
\label{eq:power}
\end{equation}
where the bin $\mathcal S_b$ contains the $N_b$ Fourier modes with magnitudes in the spherical shell $b\,k_F\leq|\mathbf{k}|<(b+1)k_F$ of width $k_F=2\pi/L_{\rm box}$, and $k_b=N_b^{-1}\sum_{\mathbf{k}\in\mathcal S_b}|\mathbf{k}|$ is their mean wavenumber.
Applying Eq.~\eqref{eq:power} to the reconstruction~$\hat\delta$ and the target~$\delta$, we define the cross-correlation coefficient
\begin{equation}
r(k)=\frac{P_{\hat\delta\delta}(k)}
{\sqrt{P_{\hat\delta\hat\delta}(k)P_{\delta\delta}(k)}},
\end{equation}
in which the~$W_2$ factors cancel.

\paragraph{Bispectrum.}
We bin Fourier modes into spherical shells~$b$ of width~$k_F$ centered at~$k_b=b\,k_F$ ($b=1,2,\dots$), so that a shell collects the modes with~$|\,|\mathbf k|-k_b|<k_F/2$.
For each shell we construct the band-limited real-space field~$\delta_{w,b}(\mathbf x)$ by inverse-transforming~$\widehat\delta_w(\mathbf k)$ with all modes outside the shell set to zero, and use the standard FFT estimator~\cite{scoccimarro:2000,Watkinson:2017kfg,Tomlinson:2019bjx}
\begin{equation}
\widehat B(k_1,k_2,k_3)=\frac{V^2}{N^3N_{\rm tri}\,W_3}
\sum_{\mathbf x}\delta_{w,b_1}(\mathbf x)
\delta_{w,b_2}(\mathbf x)\delta_{w,b_3}(\mathbf x),
\label{eq:bispectrum}
\end{equation}
where~$N_{\rm tri}$ counts closed mode triplets~$\mathbf k_1+\mathbf k_2+\mathbf k_3=0$.
We retain ordered closed shell triples with~$k_1\ge k_2\ge k_3$. We further select triangles with each side satisfying~$k\lesssim 0.77\,h/\mathrm{Mpc}$, close to two-thirds of the grid Nyquist scale, which guarantees that the real-space triple product of Eq.~\eqref{eq:bispectrum} picks up no spurious, aliased triangle closures.

The estimator defined in Eq.~\eqref{eq:bispectrum} averages over all triangle orientations.
To resolve radial long-mode correlations, we measure the alignment of a mode with the line of sight by~$\mu=k_\parallel/k$ and split the range~$\mu\in[0,1]$ into five equal-width bins with index~$a=1,\dots,5$ (with~$a=5$ collecting the most radially aligned modes,~$\mu\ge0.8$).
The band-limited field~$\delta_{w,b,a}(\mathbf x)$ is then built from the shell modes~$b$ within the alignment bin~$a$.
Replacing~$\delta_{w,b_i}\to\delta_{w,b_i,a_i}$ in Eq.~\eqref{eq:bispectrum} then defines the bispectrum~$\widehat B_c(k_1,k_2,k_3)$ of one orientation cell~$c=(a_1,a_2,a_3)$, normalized by the mode triplets~$N_{{\rm tri},c}$ of that cell.
Summing the cells with~$N_{{\rm tri},c}$ weights recovers the isotropic estimator.
A directional selection retains a subset~$\mathcal S$ of orientation cells~$c$ (e.g.~$\mu\ge0.8$ on the smallest-$k$ mode in Fig.~\ref{fig:los_bispectrum}).
For each triangle configuration and simulation~$s$, we then compute the mode-count-weighted average over the selected orientation cells,
\begin{equation}
\widehat B_{\mathcal S}^{(s)}(k_1,k_2,k_3)=
\frac{\sum_{c\in\mathcal S}N_{{\rm tri},c}\,\widehat B^{(s)}_c(k_1,k_2,k_3)}
{\sum_{c\in\mathcal S}N_{{\rm tri},c}},
\label{eq:bsel}
\end{equation}
estimating the bispectrum over the union of the retained cells.
Since the same window is applied to both fields, the ratio and cross-correlation statistics are only mildly sensitive to the window choice.

\section{Non-periodic global-receptive-field benchmark}
\label{app:toy}
We first directly compare long-horizon reconstruction capabilities and the temporal receptive fields of the two architectures using a synthetic benchmark.
We build paired realizations of scalar ``videos'' on a~$[256,64,64]$ grid, driven by the same per-slice Gaussian
random field 
$\varepsilon_t(x,y)$ with
transverse spectrum~$P(k_\perp)\propto\exp[-(k_\perp/k_0)^4]$, unit variance and zero mean.
The fields are defined by the recursive relations
\begin{equation}
A_t = a\,A_{t-1}+\sqrt{1-a^2}\,\varepsilon_t,
\qquad
B_t = b\,B_{t-1}+\sqrt{1-b^2}\,\varepsilon_t,
\qquad
A_0=B_0=\varepsilon_0 ,
\label{eq:toy_process}
\end{equation}
with the recursion acting independently at every transverse pixel and each frame being unit-variance by construction.
Since no transverse context exists by construction, reconstructing~$B$ from~$A$ forces a model to integrate the input over the temporal axis.
We fix~$a=0.8$ so the input field~$A$ loses memory of its past states within a few slices.
We tune~$b$ to control the long-range correlation of the target field across frames, parameterizing it by the frame separation~$\tau$ defined in terms of the temporal autocorrelation~$\mathrm{Corr}(B_t,B_{t+\tau})^2=b^{2\tau}=0.05$.
We sweep~$\tau\in\{32,64,128,256\}$ on~$256\times64^2$ videos and compare a \vit{} with \unet{} models of~$L=2,3,4$ resolution levels.

\begin{figure}
\centering
\begin{subfigure}[b]{0.44\textwidth}
  \centering
  \includegraphics[width=\textwidth]{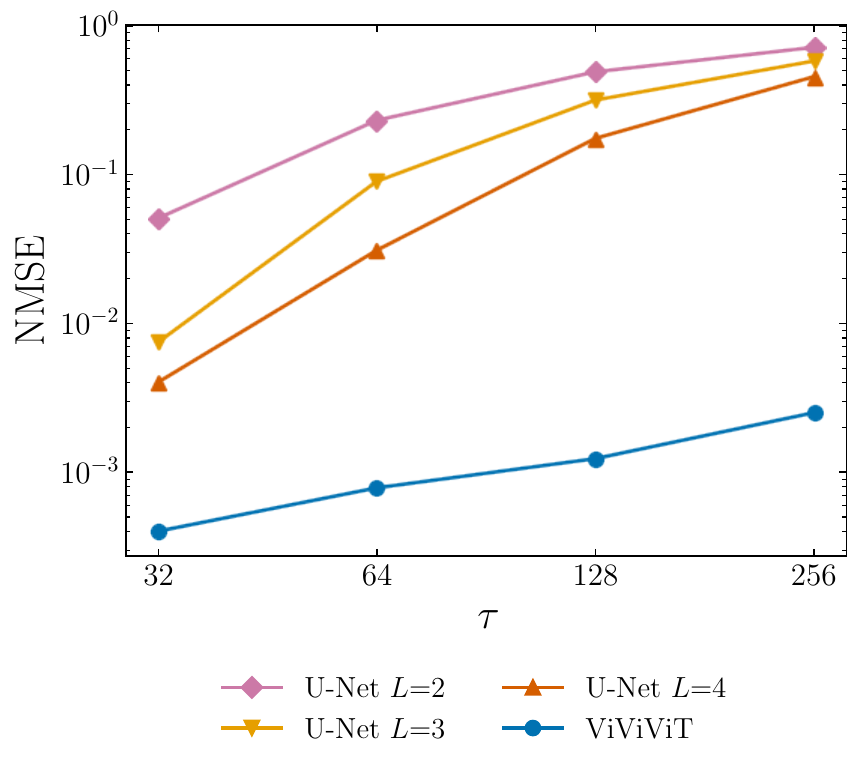}
  \caption{Error versus target memory}
  \label{fig:toy_curve}
\end{subfigure}
\hfill
\begin{subfigure}[b]{0.55\textwidth}
  \centering
  \includegraphics[width=\textwidth]{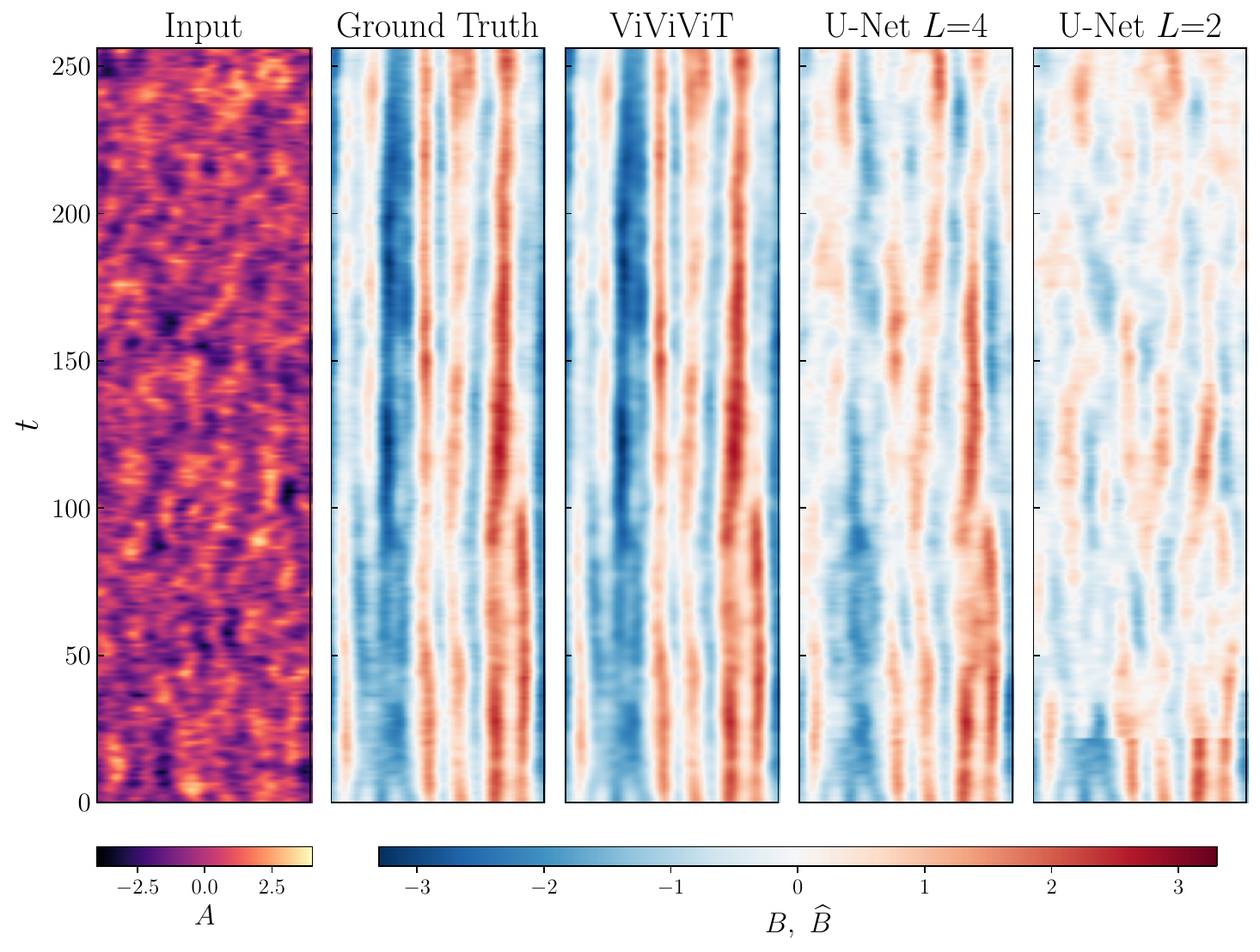}
  \caption{Representative longitudinal slices at~$\tau=256$}
  \label{fig:toy_slices}
\end{subfigure}
\caption{\justifying
\textbf{Global temporal receptive field.}
A bounded receptive field caps reconstruction quality once the target's memory exceeds it, while the global temporal reach of \texttt{ViViViT} does not degrade.
(a) Normalized mean-squared error, Eq.~\eqref{eq:nmse}, pooled over 50 held-out realizations and averaged over the 64 central slices.
\vit{} stays near~$10^{-3}$ as the memory reaches the full sequence; each bounded \unet{} degrades as~$\tau$ exceeds its usable reach.
(b) The input~$A$ decorrelates within a few slices, while the ground-truth target~$B$ carries coherent structure spanning the whole axis. \texttt{ViViViT} reproduces that slice-spanning structure; the best-performing \unet{} ($L=4$) recovers only its short-range component.}
\label{fig:toy}
\end{figure}

Since the input is an invertible filter of the same Gaussians that drive the target, a model that sees the entire input history can potentially reconstruct the target exactly.
We quantify reconstruction quality by the normalized mean squared error,
\begin{equation}
\mathrm{NMSE}(t)=
\frac{\sum_{\mathrm{sims}}\sum_{x,y}\bigl[\widehat{B}_t(x,y)-B_t(x,y)\bigr]^2}
     {\sum_{\mathrm{sims}}\sum_{x,y}\,B_t(x,y)^2},
\label{eq:nmse}
\end{equation}
evaluated per deep-axis slice~$t$ over the 50 held-out realizations.
In Fig.~\ref{fig:toy} we show the NMSE averaged over the 64 central slices and over test simulations. 
The 3D \unet{} variants, due to their finite receptive fields, show increasing reconstruction error as the target's memory grows. Each added resolution level extends the usable correlation length, but the reconstruction error follows the same trend, and by~$\tau=256$ even the deepest \unet{} misses the target almost entirely ($\mathrm{NMSE}\simeq 0.44$).
In contrast, \texttt{ViViViT} reconstructs the target at the~$10^{-3}$ level at every~$\tau$.
The reconstructed slices in the right panel of Fig.~\ref{fig:toy} visually show this result for the~$\tau=256$ case: the deepest \unet{} recovers the short-range component of the target, but loses the structure spanning the temporal axis, which the transformer reproduces.
When a target is coupled to the observable through correlations longer than the convolutional receptive field, a bounded-reach model has an intrinsic reconstruction limit that a \texttt{ViViViT} can lift through a global temporal reach.
In this work, we are using factorized attention to achieve both short and long range couplings, but various time-aware architectures can be explored to efficiently encode such multi-range correlations. 

The cosmological task of Section \ref{sec:results} is different in this respect.
In a lightcone, distinct spatial regions at distinct epochs share a causal history
that connects the evolution of the observed field.
The brightness temperature~$\delta T_{21}$ is foremost an equal-time tracer of~$\delta_m$ at a given redshift, so density reconstruction does not require global temporal reach by itself; intrinsically redshift-dependent signals, such as exotic energy injection or radiative couplings that accumulate along the lightcone, are the applications for which this capability may become necessary and where a bounded-reach \unet{} may struggle to recover the full signal.
This benchmark demonstrates that our transformer can exploit long-range temporal structure when the reconstruction task requires it.

\bibliographystyle{unsrtnat}
\bibliography{bibo,bibo_recast_additions,collaborator_additions}

\end{document}